\documentclass[useAMS,usenatbib]{mn2e}
\usepackage{graphicx, txfonts}
\usepackage{url}
\usepackage{multirow}

\usepackage[usenames,dvipsnames]{color}

\newcommand{\AlII}{Al\,\textsc{ii}}

\newcommand{\CI}{C\,\textsc{i}}
\newcommand{\CII}{C\,\textsc{ii}}

\newcommand{\DI}{\textrm{D}\,\textsc{i}}
\newcommand{\FeII}{Fe\,\textsc{ii}}

\newcommand{\HI}{\textrm{H}\,\textsc{i}}

\newcommand{\Lya}{Ly$\alpha$}

\newcommand{\NHI}{$N(\textrm{H}\,\textsc{i})$}

\newcommand{\NI}{N\,\textsc{i}}

\newcommand{\OI}{O\,\textsc{i}}

\newcommand{\SII}{S\,\textsc{ii}}
\newcommand{\SiII}{Si\,\textsc{ii}}

\title[A new candidate CEMP DLA]
{A new candidate for probing Population III nucleosynthesis with carbon-enhanced damped Lyman-\boldmath{$\alpha$} systems\thanks{
Based on observations collected at the 
W.M. Keck Observatory, which is operated as a scientific partnership
among the California Institute of Technology, the University of California
and the National Aeronautics and Space Administration. The Observatory was
made possible by the generous financial support of the W.~M.~Keck Foundation.}}

\author[Cooke, Pettini and Murphy]{Ryan Cooke$^{1,2}$\thanks{email: rcooke@ast.cam.ac.uk}, 
Max Pettini$^{1,3}$,
Michael T. Murphy$^{2}$
\\
$^1$Institute of Astronomy, Madingley Road, Cambridge, CB3 0HA\\
$^2$Centre for Astrophysics and Supercomputing, Swinburne
University of Technology, Hawthorn, Victoria 3122, Australia\\
$^3$International Centre for Radio Astronomy Research,
University of Western Australia, 7 Fairway, Crawley WA 6009,
Australia
\\
}

\begin{document}

\date{Accepted . Received ; in original form }
\pagerange{\pageref{firstpage}--\pageref{lastpage}} 
\pubyear{2011}

\maketitle

\label{firstpage}

\begin{abstract}
We report the identification of a very
metal-poor damped \Lya\ system (DLA) 
at $z_{\rm abs}=3.067295$
that is modestly carbon-enhanced, 
with an iron abundance of
$\sim 1/700$ solar ([Fe/H]\,$= -2.84$)
and [C,O/Fe] $\simeq +0.6$.
Such an abundance pattern is likely to be the result of 
nucleosynthesis by massive stars.
On the basis of 17 metal absorption lines, we derive a $2\sigma$ 
upper limit on the DLA's kinetic temperature of $T_{\rm DLA}\le4700$\,K,
which is broadly consistent with the range of spin temperature
estimates for DLAs at this redshift and metallicity.
While the best-fitting abundance pattern shows the expected hallmarks 
of Population III nucleosynthesis, models of high-mass Population II 
stars can match the abundance pattern almost as well. We discuss current 
limitations in distinguishing between these two scenarios and the marked 
improvement in identifying the remnants of Population III stars 
expected from the forthcoming generation of 30-metre class telescopes.
\end{abstract}


\section{Introduction}

With the advent of the Sloan Digital Sky
Survey in combination with $8-10$ metre
class telescopes, it has become possible to 
routinely identify clouds of gas in 
the high redshift Universe that harbour 
the metals from some of the earliest 
stellar populations
\citep{Pet08a,Pen10,Bec11,Coo11b}.
The challenge we are now faced with
is to measure the allowable ranges of mass, 
explosion energy, and metallicity of these
early stars, using metal-line diagnostics
alone. In the last decade, attention has been
drawn to the diagnostic power of carbon
for probing the nature of zero or near-zero
metallicity stars. In particular, it is
believed that metal-free stars are more
likely to yield an abundance pattern
that is enhanced in carbon relative to iron
\citep{UmeNom03}. 
In fact, such signatures may be observed among
the population of metal-poor stars in the halo of
our Galaxy that exhibit a marked carbon-enhancement
in their atmosphere; these stars are known as carbon-enhanced
metal-poor (CEMP) stars, and are currently defined
to have
[C/Fe] $>+0.7$ 
(\citealt{Aok07}, although some authors adopt
[C/Fe] $>+0.5$ or $>+1.0$ as the defining
cut).\footnote{We adopt the standard notation:
[A/B] $\equiv \log(N_{\rm A}/N_{\rm B}) -
\log(N_{\rm A}/N_{\rm B})_{\odot}$, 
where $N_{\rm A, B}$ refers to the number of 
atoms of element A and B, and the second 
term refers to the solar ratio.}

Several possibilities have been put forward to 
explain the peculiar abundance patterns of these stars.
One option is that they result from
mass transfer from a now extinct asymptotic 
giant branch companion star. This type of
CEMP star also exhibits enhancements in its
s-process neutron-capture elements, and is thus 
labelled a CEMP-s star. The above mechanism is 
now a well-established means to produce a
CEMP-s star \citep{Luc05}. 
Alternatively, the carbon enhancement may be
the residual signature from a previous 
generation of stars --- possibly Population III --- 
that seeded (with high carbon abundance)
the birth cloud of the star we see today
\citep{Rya05}. 
These are known as CEMP-no stars, as they show no 
strong enhancement in their neutron-capture elements. 
Perhaps the strongest empirical 
evidence in support of the latter picture is 
the observation of an increasing fraction of 
CEMP stars with decreasing metallicity \citep{BeeChr05}. 
Furthermore, the fraction of CEMP stars that are
labelled as CEMP-no stars
becomes relatively more common at the lowest
metallicities \citep{Aok07}.

Additional evidence in support of this picture
has recently been provided by the discovery 
of a CEMP damped \Lya\ system (DLA)
at $z_{\rm abs} = 2.340$ with [Fe/H]\,$\simeq -3$
and [C/Fe]\,$= +1.53$ \citep{Coo11a}. 
These authors proposed that such DLAs could be the 
missing link between 
the enrichment of a primordial cloud
by Population III stars, and the later 
incorporation of the trace metals into objects such as the 
CEMP-no stars seen in the halo of our Galaxy. 
Indeed, models of metal-free nucleosynthesis
are able to successfully reproduce the observed
abundance pattern for this DLA \citep{KobTomNom11}.
However, there may be an additional contribution to the metal
content of this DLA from long-lived second generation asymptotic
giant branch stars \citep{SalFer12}, which are known to
be prolific carbon producers.

\begin{figure*}
  \centering
  \includegraphics[angle=0,width=140mm]{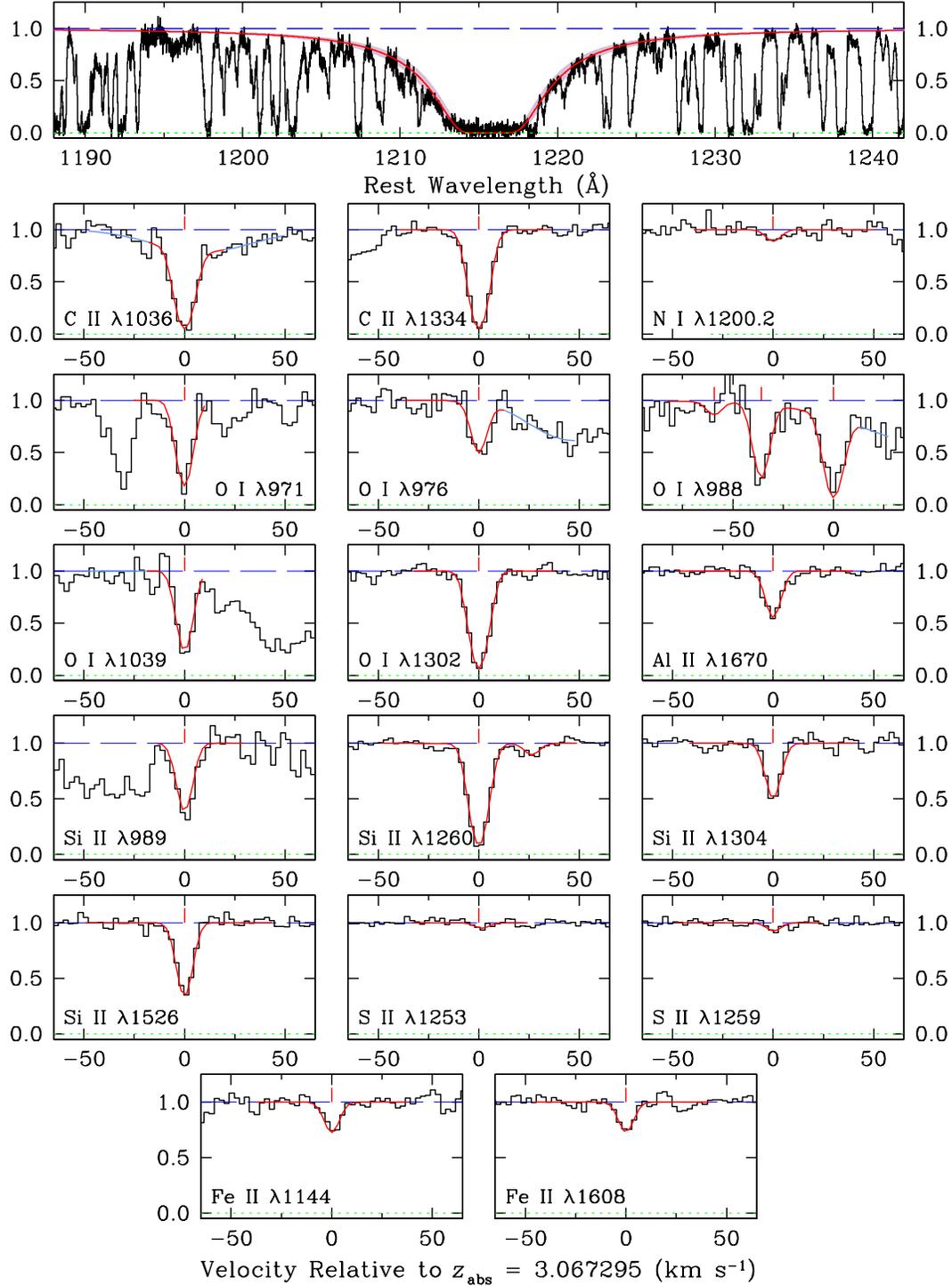}
  \caption{
Selected absorption lines in the HIRES spectrum of the
$z_{\rm abs}=3.067295$ DLA towards the 
QSO J1358$+$6522.
Black histograms are the data, and red continuous lines
show the best fitting theoretical profiles computed
with \textsc{vpfit} as discussed in Section~\ref{sec:line_fitting}.
The top panel shows the \HI\ \Lya\ absorption line
together the damped profile corresponding 
to a column density $\log$\,[\NHI/cm$^{-2}] = 20.47\pm0.07$. 
Blue lines in some of the panels are used to indicate 
unrelated \Lya\ forest blends. The weak feature
at $+26$\,km\,s$^{-1}$ in the \SiII\,$\lambda 1260.4221$ 
panel is  \FeII\,$\lambda 1260.533$.
The $y$-scale is residual intensity.
}
  \label{fig:montage}
\end{figure*}

\begin{table*}
\centering
\caption{
\textsc{Ion column densities of the DLA towards J1358$+$6522 at $z_{\rm abs}=3.067295$}}
    \begin{tabular}{@{}lp{1.8in}ccccc}
    \hline
  \multicolumn{1}{l}{Ion}
& \multicolumn{1}{c}{Transitions used}
& \multicolumn{1}{c}{log $N$(X)/${\rm cm}^{-2}$}
& \multicolumn{1}{c}{log $\epsilon$(X)$^{a,\,b}_{\rm DLA}$}
& \multicolumn{1}{c}{log $\epsilon$(X)$^{a,\,c}_{\odot}$}
& \multicolumn{1}{c}{[X/H]$^{b,\,d}$}
& \multicolumn{1}{c}{[X/Fe]}\\
    \hline
\HI    &  1025, 1215                          &  20.47 $\pm$ 0.07 &  ---  &  12.0  &  ---  &  ---  \\
\CII   &  1036, 1334                          &  14.65 $\pm$ 0.10 &  $6.18\pm0.10$  &  8.43  &  $-2.25\pm0.10$  &  $+0.59\pm0.10$  \\
\NI    &  1200.2                              &  12.62 $\pm$ 0.14 &  $4.15\pm0.14$  &  7.83  &  $-3.68\pm0.14$  &  $-0.84\pm0.14$  \\
\OI    &  971, 976, 988$\times3$, 1039, 1302  &  14.94 $\pm$ 0.05 &  $6.47\pm0.05$  &  8.69  &  $-2.22\pm0.05$  &  $+0.62\pm0.06$  \\
\AlII  &  1670                                &  11.92 $\pm$ 0.03 &  $3.45\pm0.03$  &  6.44  &  $-2.99\pm0.03$  &  $-0.15\pm0.04$  \\
\SiII  &  989, 1260, 1304, 1526               &  13.40 $\pm$ 0.03 &  $4.93\pm0.03$  &  7.51  &  $-2.58\pm0.03$  &  $+0.26\pm0.04$  \\
\SII   &  1253, 1259                          &  13.11 $\pm$ 0.09 &  $4.64\pm0.09$  &  7.14  &  $-2.50\pm0.09$  &  $+0.34\pm0.09$  \\
\FeII  &  1144, 1260, 1608                    &  13.10 $\pm$ 0.03 &  $4.63\pm0.03$  &  7.47  &  $-2.84\pm0.03$  &  $+0.00$  \\
    \hline
    \end{tabular}
\begin{flushleft}
$^{\rm a}${$\log \epsilon{\rm (X)}=12+\log N({\rm X})/N({\rm H})$}.
\\
$^{\rm b}${We have not included the uncertainty in \HI}.
\\
$^{\rm c}${\citet{Asp09}}.
\\
$^{\rm d}${[X/H]$_{\rm DLA} \equiv \log \epsilon{\rm (X)}_{\rm DLA} - \log \epsilon{\rm (X)}_{\odot
}$}.\\
\end{flushleft}
\label{tab:coldens}
\end{table*}

DLAs are clouds of neutral gas
seen in absorption 
against the light of background QSOs;
their properties have been reviewed by 
\citet{Wol05}. As discussed by
\citet{Pet04} and more recently
\citet{Coo11b}, measurements of their chemical composition
complement very effectively local stellar abundance studies.
While most DLAs are metal-poor, with typical
metallicities of $\sim 1/30$ of solar
\citep{Ell12}, interest has recently
focused on identifying and
studying the most metal-poor DLAs, 
with [Fe/H]\,$ < -2$, which may still retain
the nucleosynthetic signatures of the first episodes
of metal enrichment in the Universe
\citep{Ern06, Pet08a, Pen10, Coo11b}.

Empirically, it is found that 
the most metal-poor DLAs tend to
have very quiescent kinematics,
with most of the metal line absorption concentrated
in one or two components of low velocity 
dispersion.
For example, the $z_{\rm abs} = 2.340$
CEMP DLA reported by \citet{Coo11a}
consists of a single absorption component 
with an internal one-dimensional velocity dispersion
$\sigma = 1.7$\,km~s$^{-1}$.
In this extreme regime it becomes important
to differentiate between bulk gas motions
which broaden all absorption lines by the 
same amount, and thermal motions which,
for the same gas temperature, 
broaden lighter elements more than heavier ones
(see Section~\ref{sec:line_fitting}).
Unfortunately, with present instrumentation
it is generally not possible to fully resolve
the metal line profiles produced by
such quiescent gas, although under favourable
circumstances a  
curve-of-growth analysis 
can help to separate the relative contributions
of turbulent and thermal broadening to
the line equivalent widths
\citep[e.g.][]{Ome01, Jor09, Tum10, Car11}.

In their analysis of the CEMP DLA at 
$z_{\rm abs} = 2.340$ \citet{Coo11a}
assumed that turbulent motions dominate,
as is the case if the gas temperature is less
than a few thousand degrees. Higher
temperatures would have the effect of decreasing
selectively the carbon abundance, and hence the
degree of carbon enhancement (see Section~\ref{sec:line_fitting}
for further details), although the resulting line profiles
are a poorer fit to the data \citep{Coo11c, Car12}. 

In order to make progress on these issues it is important
to identify and study further examples of CEMP DLAs.
In this paper we report the discovery of 
a new candidate with a lower carbon enhancement than
the first example of this class, but where the gas
temperature can be constrained within a narrower range. 
In Section~\ref{sec:obs} we describe the observations 
and data reduction procedures, saving the discussion of
line profile fitting and the abundance analysis 
to Section~\ref{sec:line_fitting}. We discuss
the possibility of using such systems to measure
the kinetic temperature of DLAs in
Section~\ref{sec:gastemp}, before comparing
the best-fitting abundance pattern to
models of stellar nucleosynthesis in 
Section~\ref{sec:nucleo} and concluding 
in Section~\ref{sec:conc}.

\section{Observations and Data Reduction}
\label{sec:obs}

To search for additional candidate CEMP DLAs
out of the increasing database of damped
systems, we target metal-poor systems 
showing an
unusually strong \CII\,$\lambda 1334$
absorption line in spectra
recorded at medium resolution
(full width at half maximum,
FWHM\,$\simeq 50$--$100$\,km\,s$^{-1}$). 
While this relatively coarse resolving power
is insufficient for a reliable abundance analysis
\citep{Coo11b}, it does provide the means to
identify some of the most promising 
candidates.

We selected the DLA at $z_{\rm abs}= 3.0675$
towards the quasar J1358$+$6522 
on the basis of the Echelle Spectrograph and Imager
(ESI)  observations with FWHM $\simeq 60$\,km\,s$^{-1}$
by \citet{Pen10}. These authors reported
[O/H]\,$= -3.08 \pm 0.15$ and 
[C/O]\,$=+0.44 \pm 0.43$; 
given that DLAs in 
this metallicity regime commonly exhibit an
$\alpha$-enhancement of [O/Fe] $\simeq +0.4$ \citep{Coo11b},
the \citet{Pen10} data suggest that [C/Fe] $\simeq +0.8$.

We conducted follow-up observations of J1358$+$6522
for $28\,950$\,s divided into nine exposures
with the W.~M.~Keck Observatory's High Resolution Echelle 
Spectrometer (HIRES; \citealt{Vog94}) on 2011 March 23, 24
(program ID: A152Hb). We used the C5 decker
(a $7.0 \times 1.148$\,arcsec slit) which delivers
a nominal resolution of $8.1$\,km\,s$^{-1}$ for a 
uniformly illuminated slit 
(but see Section~\ref{sec:line_fitting}, 
where we adopt a lower FWHM 
resolution of $7.0$\,km\,s$^{-1}$ due to the 
sub-arcsecond seeing conditions), 
which we sample with $\sim 3$ pixels. 
We used the red cross-disperser with the
WG360 filter and  $2\times2$ on-chip binning.

The data were reduced with the \textsc{makee}
data reduction pipeline maintained by
Tom Barlow.\footnote{We used \textsc{makee} 
version 5.2.4, available for download from\\
\texttt{http://spider.ipac.caltech.edu/staff/tab/makee/}}
\textsc{makee} performs the usual steps relevant to
echelle data reduction, including  bias subtraction,
flat-fielding, order definition and extraction.
Each science exposure was followed 
by a ThAr hollow-cathode lamp
frame which was used to wavelength 
calibrate the extracted orders. The wavelength
scale was converted to vacuum heliocentric
using the software package
\textsc{uves\_popler}.\footnote{\textsc{uves\_popler} 
can be downloaded from\\
\texttt{http://astronomy.swin.edu.au/${\sim}$mmurphy/UVES\_popler/}} 
This software was also used to combine 
the extracted orders from all exposures, whilst
rejecting deviant pixels and unusable orders.
Finally, the combined spectrum was 
normalised by fitting and dividing 
out the quasar continuum and emission lines.
The metal absorption features of interest were 
then extracted in $\pm 200$\,km\,s$^{-1}$ windows
about the line centroid, and a fine adjustment
to the continuum was applied if necessary. The
resulting signal-to-noise ratio of the data at 
5000\,\AA\ is S/N $\sim 30$ per resolution element.
Examples of absorption lines in the $z_{\rm abs}= 3.0675$
DLA are collected in Figure~\ref{fig:montage}.

\section{Analysis and Results}
\label{sec:line_fitting}

\subsection{Profile Fitting}

As can be seen from Figure~\ref{fig:montage}
and Table~\ref{tab:coldens}, our HIRES spectrum 
covers many transitions of the dominant ionisation
stages of C, N, O, Al, Si, S and Fe, as well as lines
in the Lyman series of \HI. All the metal lines
exhibit a single, narrow absorption component
which, with FWHM\,$\simeq 5$ pixels, is just
resolved with our instrumental resolution of 
$\sim 3$ pixels.

In order to deduce the corresponding element
abundances, we modelled the absorption lines 
with theoretical Voigt profiles (convolved
with the instrumental broadening function)
using the $\chi^{2}$-minimisation software 
\textsc{vpfit}\footnote{\textsc{vpfit} is available from\\
\texttt{http://www.ast.cam.ac.uk/${\sim}$rfc/vpfit.html}},
which returns the best-fitting Doppler parameter
$b = \sqrt{2} \sigma$ and 
redshift $z_{\rm abs}$ of the DLA, in addition to the column densities $N$
of the available ions. The line fitting procedure
consisted of three main steps, which we now outline.

\subsection{Instrumental Resolution}

In the first step, we determined the true value
of the instrumental resolution which is likely to be lower
than the nominal value of FWHM\,$= 8.3$\,km~s$^{-1}$
appropriate to the uniformly illuminated slit 
provided by the C5 HIRES decker, given that the
seeing during our observations was better
than the 1.15\,arcsec width of the slit.
Examination of the emission line profiles from 
the Th-Ar hollow-cathode lamp showed that,
at the signal-to-noise ratio of our data, a 
gaussian profile is a good approximation to the
instrumental broadening function.

We assessed the resolution of our spectrum 
empirically from the metal lines in the 
$z_{\rm abs}= 3.0675$ DLA---which are 
the narrowest features available in the 
spectrum---using \textsc{vpfit} in a series
of model fits which differed by the assumed 
value of the instrumental broadening FWHM
and noting the corresponding value of $\chi^{2}$,
as described in \citet{Coo11a}.
The minimum value of $\chi^{2}$ was 
found to be that for 
an instrumental resolution
FWHM$_{\rm instr} = 7.0$\,km\,s$^{-1}$,
which we adopted in all
the subsequent model fitting.
Lowering the instrumental resolution
from the nominal 
FWHM$_{\rm instr} = 8.3$\,km\,s$^{-1}$,
to FWHM$_{\rm instr} = 7.0$\,km\,s$^{-1}$
has only a modest effect on the derived element
abundances, since for most elements we cover
at least one absorption line which is optically
thin. Thus, for example, the change in the
column density of \FeII\ is only
$\Delta \log N{\rm (Fe\,\textsc{ii})} = -0.02$.
The largest change is for \CII,
$\Delta \log N{\rm (C\,\textsc{ii})} = -0.07$,
because both \CII\,$\lambda 1334$ and 
$\lambda 1036$ transitions are moderately
saturated (see Figure~\ref{fig:montage}).

\subsection{Abundance Analysis}

Having determined the most likely value
of the instrument resolution, we used \textsc{vpfit}
to derive the absorption parameters,
or `cloud model',
that best represent the data. 
When transitions from elements of different
atomic mass are available, as is the case here,  
\textsc{vpfit} can solve separately 
for the macroscopic, or turbulent, broadening
parameter, $b_{\rm turb}$, which is independent
of atomic mass, and the microscopic, or thermal,
broadening parameter, $b_{\rm th}$, which
is related to the temperature $T$ of the gas
and the mass of the ion $m$ via the standard
relation: $b_{\rm th}^2 = 2kT/m$, where $k$ is
the Boltzmann constant. The two contributions
add in quadrature to give the total broadening:
$b_{\rm tot}^2 = b_{\rm turb}^2 + b_{\rm th}^2$.

Given the simplicity of the kinematics of the DLA
under consideration, where all the absorption lines
due to ions that are dominant in neutral gas
(in our case \OI, \NI, \CII, \AlII, \SiII, \SII\ and \FeII)
appear to consist of a single, narrow component,
we forced \textsc{vpfit}
to find the best-fitting values of $z_{\rm abs}$,
$b_{\rm turb}$ and $T$ \textit{common to all the above species}.
We found that a single component 
with redshift $z_{\rm abs}=3.067295 \pm 0.000001$, 
Doppler parameter $b_{\rm tot} =3.00 \pm 0.06$\,km~s$^{-1}$, 
and the ion column densities (and associated errors)
as listed in Table~\ref{tab:coldens}, provides 
the best representation of the data. 
The corresponding line-profile fits are overplotted
on the data in Figure~\ref{fig:montage}. 
Interestingly, \textsc{vpfit} favours low values
of temperature such that the broadening
is dominated by turbulent motions
(further details in Section~\ref{sec:thermbroad}). 

Since the species observed are the dominant
ion stages of the corresponding elements in \HI\ regions,
the element abundances can be obtained directly 
by dividing the ion column densities 
in the third column of Table~\ref{tab:coldens}
by the \HI\ column density
$\log N{\rm (H \textsc{i})} = 20.47 \pm 0.07$.
Comparison with the solar abundance scale
of \citet{Asp09} finally gives the abundance pattern
relative to solar listed in the penultimate column
of Table~\ref{tab:coldens}. 

We find [Fe/H]\,$=-2.84 \pm 0.03$,
making this DLA one of the most metal-poor known.
We also find [O/H]\,$=-2.22 \pm 0.05$
and [C/O]\,$\simeq 0$; these values
are significantly different from those
reported by \citet{Pen10} from data of lower
spectral resolution, 
[O/H]\,$=-3.08 \pm 0.15$
and [C/O]\,$\simeq +0.45$.
As discussed by \citet{Coo11b}, such
differences are not unexpected 
when comparing medium- and high-resolution
data (and are indeed the reason why high resolution
follow-up observations of candidate metal-poor DLAs
are essential for accurate abundance determinations, 
despite the substantial investment in observing time
required).

\begin{figure}
  \centering
  \includegraphics[angle=0,width=80mm]{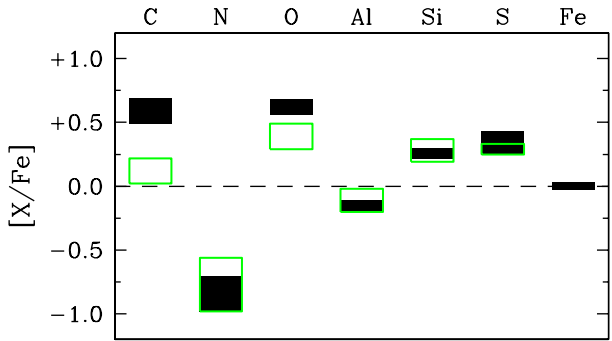}
  \caption{
The abundance pattern for the
carbon-enhanced DLA (black boxes, where the
height of each box represents the $1\sigma$
statistical uncertainty) is shown for the
best-fitting cloud model. For comparison,
we also show the abundance pattern of a
typical very metal-poor carbon-normal DLA
(green boxes, the height of each box shows
the $1\sigma$ dispersion in the population;
\citealt{Coo11b}).
The dashed line is drawn at the solar value.
  }
  \label{fig:turbpatt}
\end{figure}

In Figure~\ref{fig:turbpatt} we show the abundance
pattern of the DLA relative to Fe, together 
with that of a typical `carbon-normal' DLA, taken from the
recent survey of very metal-poor DLAs by \citet{Coo11b}.
Both C and O appear enhanced relative to Fe by larger
factors than is normally the case, although the C enhancement
is much less pronounced than in the first reported case
of a CEMP DLA, [C/Fe]\,$= +1.53$ in
the $z_{\rm abs} = 2.340$ 
DLA in line to the QSO J0035$-$0918
\citep{Coo11a}.
Before considering possible interpretations of the
chemical composition we have uncovered
(Section~\ref{sec:nucleo}),
we discuss the effect of thermal broadening
on the derived element abundances,
estimated through a third and final set
of \textsc{vpfit} simulations.

\subsection{Thermal Broadening?}
\label{sec:thermbroad}

As explained in the previous section, the 
cloud model which best fits the observed
metal line profiles is one in which thermal
motions do not make a significant contribution
to the line broadening, i.e. the gas in this DLA 
appears to be `cold'.
We deduced an upper limit
to the gas temperature by no longer treating
$b_{\rm th}$ as a free parameter in the 
line fitting procedure, but instead
forcing \textsc{vpfit} to adopt a fixed value of $T$
($b_{\rm turb}$ remained a free parameter 
throughout these tests).
In a procedure similar to that described by 
\citet[][see also \citealt{LamMarBow76}]{Car12},
we considered a grid of temperatures within the
range $100\,{\rm K} \leq T \leq 1.3 \times 10^{4}\,{\rm K}$,
and for each value of $T$, rerun \textsc{vpfit} to find the best-fitting
model parameters and corresponding value of $\chi^2$.
In these tests we excluded the weak \NI\ and \SII\ lines
(see Figure~\ref{fig:montage}), 
because they offer no constraint on the line broadening.

\begin{figure}
  \centering
  \includegraphics[angle=0,width=80mm]{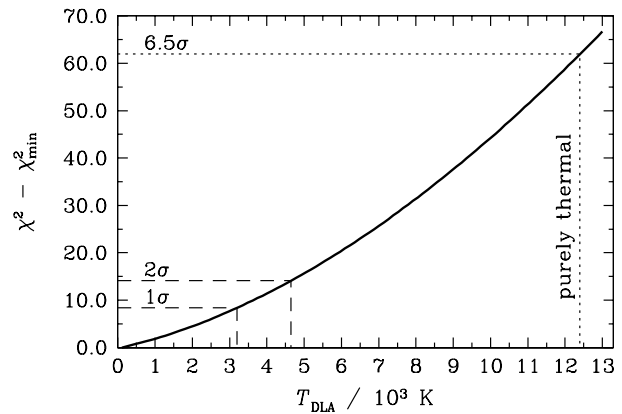}
  \caption{
On the basis of the available metal lines, we show the 
differential $\chi^{2}$ value, 
$\Delta\chi^{2}(T_{\rm DLA}) = \chi^{2}(T_{\rm DLA}) - \chi^{2}_{\rm min}$, 
for a range of plausible kinetic temperatures for 
the new candidate CEMP DLA (solid curve). We also
show the $1\sigma$ and $2\sigma$ confidence regions
for a $\chi^{2}$ distribution with $7$ degrees of
freedom (horizontal dashed lines; see text for further 
details). This suggests that the DLA's kinetic 
temperature is $T_{\rm DLA}\le4700$\,K. A purely
thermal model, indicated by the dotted lines, is 
ruled out at $6.5\sigma$.
  }
  \label{fig:chisq_temp}
\end{figure}

In Figure~\ref{fig:chisq_temp}, we plot the resulting 
differential $\chi^{2}$ value,
$\Delta\chi^{2}(T_{\rm DLA}) = \chi^{2}(T_{\rm DLA}) - \chi^{2}_{\rm min}$ 
for each temperature, where $\chi^{2}_{\rm min}$ 
refers to the $\chi^{2}$ of the best-fitting model.
Noting that \emph{seven} 
parameters can alter the total line width 
(i.e. the thermal and turbulent Doppler widths in 
addition to five column density estimates which 
are weakly anticorrelated with the total line width), 
the $70$ and $95$ percent confidence regions of the 
$\chi^{2}$ distribution (i.e. $1\sigma$ and $2\sigma$
for a normal distribution)
correspond to $\Delta\chi^{2}=8.4$ and $14.1$ 
respectively. These are indicated with dashed lines in 
Figure~\ref{fig:chisq_temp}. 
From this test we derive a $2 \sigma$ upper limit to 
the DLA's kinetic temperature of $T_{\rm DLA} \le 4700$\,K.
Furthermore, we remark that a cloud model
where the line profiles are broadened
\emph{entirely} by thermal motions 
(resulting in a temperature 
$T_{\rm DLA}= (1.24 \pm 0.06) \times10^{4}$\,K),
gives  $\Delta\chi^{2}=62.0$ and is thus
ruled out at more than $6.5\sigma$.

The effect of raising the gas temperature
from the best-fitting low values to the 
$2\sigma$ upper limit is most pronounced 
for the \CII\ column density and corresponding
C abundance. The reasons are two-fold.
First, the two \CII\ lines 
accessible, $\lambda 1334$ and $\lambda 1036$,
are both strong and close to saturation 
(see Figure~\ref{fig:montage}). 
For all other species transitions 
are available which are so weak that
the column density is
almost independent of the Doppler parameter,
and is thus insensitive to the DLA temperature.
Second,  C is the lightest among the metals observed,
and is thus the one with the largest value of $b_{\rm th}$
at a given temperature.
Consequently, adopting the $2 \sigma$
upper limit $T_{\rm DLA} \le 4700$\,K
has the effect of \textit{reducing}  
[C/Fe]  by 0.37\,dex relative to the value
in Table~\ref{tab:coldens}, diluting
the degree of C enhancement deduced for this DLA.
On the other hand,  [O/Fe] is unchanged
(for the reasons just explained: 
both O and Fe have weak absorption
lines in our spectrum); thus, even 
for the upper limit on the temperature,
the [O/Fe] abundance in this DLA
 is $\sim 2\sigma$ away from the 
values typical of the metal-poor DLA population
(Figure~\ref{fig:turbpatt}).

The usefulness of the analysis described above
is limited by the fact that the masses of the elements 
included in our data set span less
than a factor of five between C and Fe, so that
the corresponding values of $b_{\rm th}$ for
a given $T$ differ by little more than a factor of two
(recall that $b_{\rm th} \propto m^{-1/2}$).
Further progress could be made by observing a
much lighter element, such as D. However, absorption lines
from the Lyman series of \DI\ can only
be resolved from the neighbouring \HI\ transitions
in high order Lyman lines \citep[e.g.][]{Ome06, Pet08b};
recording these transitions with the required S/N ratio
will involve dedicated and time consuming observations
in the ground-based UV portion of the spectrum.

In conclusion, while the most likely interpretation
(i.e. the one producing the lowest value
of $\chi^2$) of the absorption lines in the 
DLA considered here implies a C enhancement
by a factor of $\sim 4$ ([C/Fe]\,$= +0.59$),
the uncertainties 
introduced by the
range of gas temperatures allowed by the data 
lead us to consider the $z_{\rm abs} = 3.067$
DLA in front of J1358$+$6522 as a 
\emph{candidate} CEMP DLA.


\section{Gas temperature estimates in DLAs}
\label{sec:gastemp}

At present, there have been very few measurements
of the kinetic temperature for high redshift DLAs.
This is hardly surprising, since the contribution to
the total metal-line broadening in a gas with a
temperature $\lesssim10^{4}$ K is nearly undetectable,
even for systems with a relatively low velocity
dispersion ($\sigma\sim5$\,km\,\,s$^{-1}$). Indeed,
the sole direct measurement of a DLA's kinetic
temperature, on the basis of metal-lines alone,
was reported only recently by \citet{Car12},
who measured a gas temperature of
$T_{\rm DLA}=(1.2\pm0.3)\times10^{4}$\,K for
the $z_{\rm abs}=2.076$ DLA towards Q$2206-199$.

For the DLA we report herein,
the upper limit on the kinetic temperature of
$T_{\rm DLA}\le4700$\,K
is consistent with the gas residing in 
either a cold (a few $\times10^{2}$ K) or a
warm (a few $\times10^{3}$ K) neutral medium.
In fact, the absence of molecular absorption
lines (or species such as \CI) might suggest
that the DLA's gas is predominantly
warm, with a temperature that is likely
$\gtrsim3000$\,K \citep{PetSriLed00}.
As a side note, we also comment that our
derived limit for this system is consistent
with the best-fitting temperature of $4000$\,K
for the CEMP DLA towards J$0035-0918$, using
the metal absorption lines alone.

At present, the
majority of DLA temperature estimates 
rely on measuring the spin temperature from
the \HI\ $\lambda$21\,cm line and assuming that
it is thermalised \citep[e.g.][]{Kan09,Sri12}.
This may not be a valid assumption in the metal-poor
regime, where there are fewer atomic
coolants available to lower the temperature
of the gas below $T_{\rm DLA}\lesssim1000$\,K;
for temperatures above $\sim1000$\,K, collisions
are less efficient at thermalising the
\HI\ $\lambda$21\,cm line \citep[see e.g.][]{Lis01},
resulting in a spin temperature that is somewhat
less than the kinetic temperature. Although this
remains to be tested in a system where both the
spin and kinetic temperature can be derived,
the current upper limits based on line-profile
techniques are consistent with the lower
limits afforded by spin temperatures for
[Fe/H]$<-2.0$.

Certainly, the simplicity of the absorption lines
from such quiescent DLAs highlights them as
ideal systems to directly measure the kinetic
temperature of neutral gas in the metal-poor regime.
The ability to measure or place limits on the
kinetic temperature of DLAs is an exciting
prospect for future research; one could in
principle test how the gas temperature varies
as a function of metallicity, \HI\ column
density, and redshift.

\section{Comparison with nucleosynthetic yields}
\label{sec:nucleo}

We now compare the best-fitting abundance 
pattern shown in Figure~\ref{fig:turbpatt}
to illustrative models of stellar
nucleosynthesis. The ultimate aim
of this work is to decipher the  
clues into the nature of the
stars responsible
for the metal enrichment 
of the most metal-poor DLAs;
here we look critically at what 
advances can be made towards this goal
using the new data presented here.
 
At very low metallicities it is likely that
the dominant metal yields are from
massive stars ($10$--$100\,M_{\odot}$) that
explode as core-collapse supernovae (CCSNe).
At present, however, we still lack a full
physical understanding of the explosion 
mechanism that operates during the 
collapse of a massive star
(see e.g. \citealt{Mez05} 
and \citealt{OttOCoDas11} 
for recent reviews on this topic).
Despite these uncertainties, several groups
have developed theoretical models that 
parameterise the unknown physics, such as 
the degree of mixing between the stellar 
layers and the final kinetic energy of 
the explosion, in order to estimate 
the resulting nucleosynthetic yields. 
Here we consider two CCSNe model suites 
to compare to the measured abundance
pattern of the candidate CEMP DLA:
(1) the recent calculations for metal-free
Population III stars in the mass range
$10$--$100\,$M$_{\odot}$  by
\citet{HegWoo10}; and (2) the Population III/II/I
calculations by \citet{ChiLim04} for
stars with masses in the range $13$--$35\,$M$_{\odot}$.

\begin{figure*}
  \centering
 {\hspace{-0.25cm} \includegraphics[angle=0,width=80mm]{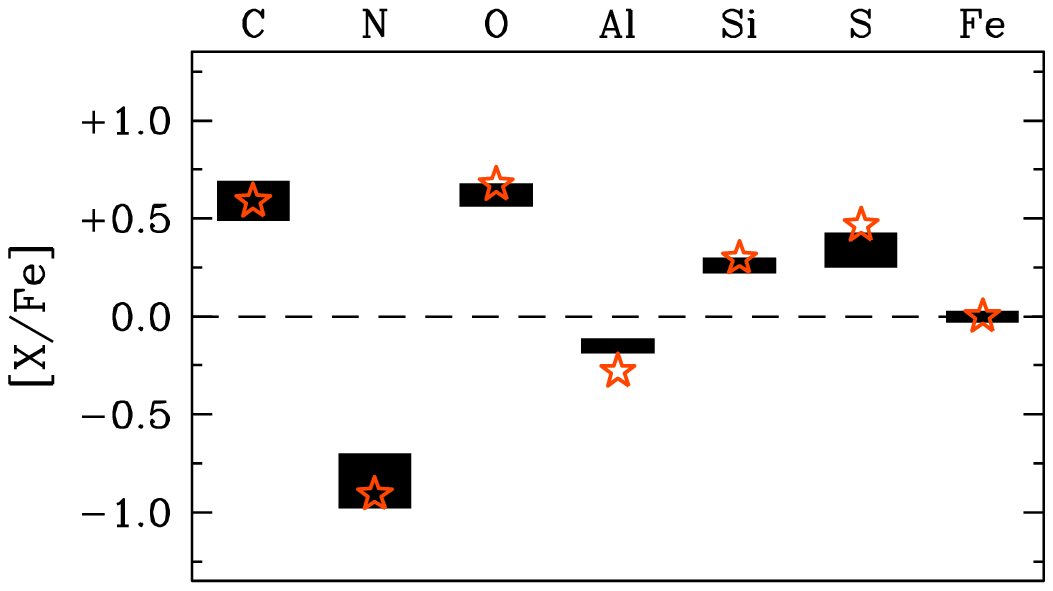}}
 {\hspace{0.25cm} \includegraphics[angle=0,width=80mm]{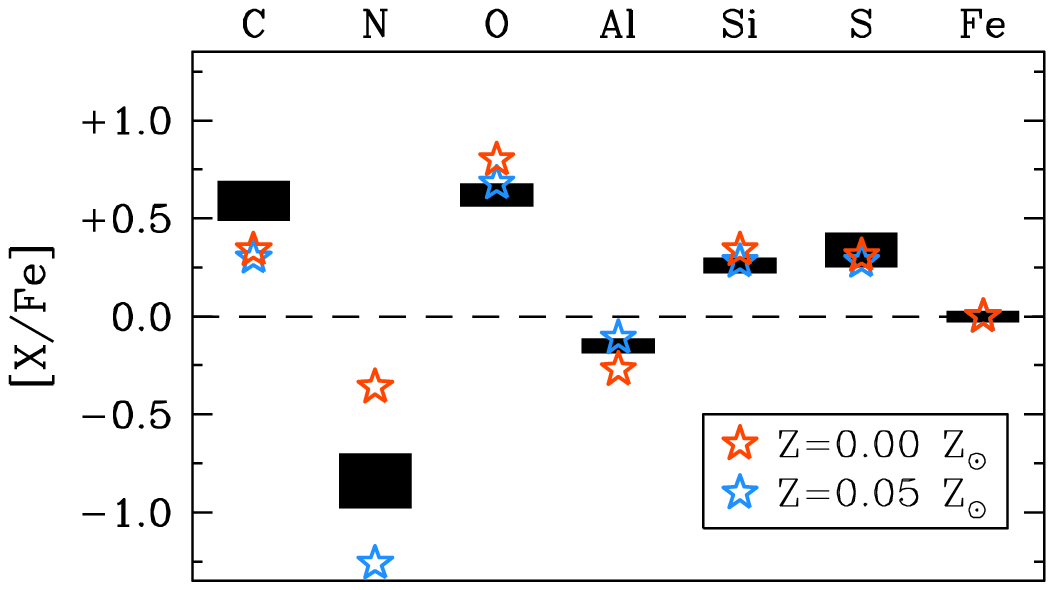}}\\
  \caption{
For the most probable abundance pattern 
(see Figure~\ref{fig:turbpatt}), we overplot the
best-fitting models of Population III (left panel stars;
\citealt{HegWoo10})
and Population II (right panel blue stars;
\citealt{ChiLim04}) nucleosynthesis. We also
plot the best-fitting Population III model from
the recent \citet{LimChi12} yield calculations
(right panel orange stars).
Statistically, the metal-free \citet{HegWoo10} model
and the $Z=0.05\,\,Z_{\odot}$ \citet{ChiLim04} model
provide an equally good fit to the data,
given that the former contains more model
parameters. The recent metal-free calculations
by \citet{LimChi12}, on the other hand, are less favoured.
Given the (unknown) model uncertainties, we consider the
data to be equally well represented by all three models.
  }
  \label{fig:nucmod}
\end{figure*}

\begin{table}
\centering
\caption{
\textsc{Summary of best-fitting model yields}}
    \begin{tabular}{@{}lccccc}
    \hline
  \multicolumn{1}{@{}l}{Model$^{a}$}
& \multicolumn{1}{c}{$Z$ ($Z_{\odot}$)}
& \multicolumn{1}{c}{Mass (M$_{\odot}$)}
& \multicolumn{1}{c}{[C/Fe]}
& \multicolumn{1}{c}{[O/Fe]}
& \multicolumn{1}{c}{Fe Mass (M$_{\odot}$)}\\
    \hline
HW10   &   0.00  &   $17^{+7}_{-5}$  & +0.60 & +0.67 & $0.017^{b}$ \\
CL04   &   0.05  &   $35$            & +0.30 & +0.68 & 0.215 \\
LC12   &   0.00  &   $35$            & +0.34 & +0.80 & 0.144 \\
    \hline
    \end{tabular}
\begin{flushleft}
$^{\rm a}${HW10: \citet{HegWoo10};
CL04: \citet{ChiLim04};
LC12: \citet{LimChi12}}.
\\
$^{\rm b}${For the HW10 model,
this is the IMF-weighted ejected Fe mass}.
\\
\end{flushleft}
\label{tab:modpars}
\end{table}

The recent database of zero-metallicity 
nucleosynthesis calculations
by \citet{HegWoo10} contains
a suite of $1440$ models, with two parameters
describing the physics of the explosion and 
one parameter describing the mixing. Although
these calculations use a fine mass-resolution
of $\sim0.1$ M$_{\odot}$, the model stars that
differ in mass by just $\sim0.1$ M$_{\odot}$ often
produce notable differences in the final
nucleosynthetic yields. We have therefore
opted to integrate the yields over a Gaussian-like
initial mass function (IMF; two additional model
parameters are used to define the centroid
[M$_{\rm IMF}$] and width [$\sigma_{\rm IMF}$] of
the IMF). In this way, we hope to obtain an approximate
handle on the mass range of the stars responsible for
the DLA's enrichment. Thus, the resulting parameter
space contains $196\,000$ combinations of the five
adjustable parameters.

To determine the best-fitting model, we use the 
\textsc{star\_fit} software provided by 
\citet{HegWoo10}.\footnote{\textsc{star\_fit} is
available for download from\\
\texttt{http://homepages.spa.umn.edu/${\sim}$alex/znuc/}}
The combination of parameters
which best fits the DLA element abundances 
has an IMF
centred on M$_{\rm IMF}=17\,$M$_{\odot}$, with a width
of $\log \sigma_{\rm IMF}/$M$_{\odot}=0.15$
(details are summarised in Table~\ref{tab:modpars}).
This model has a relatively low degree
of mixing between the stellar layers 
(approximated by a running boxcar filter with
a width equal to 2.5\% of the total 
He core mass), and a final kinetic energy 
of the ejected material of 
$1.4\times10^{51}$\,erg\,s$^{-1}$.
For this IMF-weighted model,
the average star will eject $0.017$ M$_{\odot}$ of Fe.
An Fe yield of just $0.017$ M$_{\odot}$ is quite
low for such an explosion energy,
this occurs as a result of the low degree
of mixing between the stellar layers during
the explosion.

The abundance pattern generated with this 
model is shown in the left
panel of Figure~\ref{fig:nucmod}.
Overall, there is good agreement
between calculated and measured abundances,
except possibly for Al which deviates by
$\sim 3\sigma$. 
However, the discrepancy
is only 0.13\,dex, which is small
considering the likely uncertainties
in the calculations of model yields.

We next consider the nucleosynthesis calculations 
of Population~III/II/I stars by \citet{ChiLim04}. This
database contains a set of six stars of different 
mass, spanning a range of metallicity (a further
six choices of metallicity are available, including:
$Z=0.0, 5\times10^{-5}, 0.005, 0.05, 0.3,$ and $1.0\,\,Z_{\odot}$),
and an effectively continuous range of `mass cut', which
is a parameterisation for the amount of material
that escapes the binding energy of the star 
following the explosion; this 
leads to a model with only three free
parameters. The model that comes closest
to reproducing the chemical composition of our
candidate CEMP DLA is for a
$35\,$M$_{\odot}$ star with a 
metallicity of $Z=0.05\,Z_{\odot}$
ejecting $0.206\,$M$_{\odot}$ of $^{56}$Ni
(the default value considered by 
\citealt{ChiLim04} is $0.1\,$M$_{\odot}$ of $^{56}$Ni).
A summary of the model details is presented
in Table~\ref{tab:modpars}.
In order to eject $0.2\,$M$_{\odot}$ of $^{56}$Ni,
the kinetic energy of the ejecta for a $35\,$M$_{\odot}$
star would need to be an order of magnitude larger than
a typical supernova, and will thus reside on the
hypernova branch (see e.g. Figure~1 from \citealt{TomUmeNom07}).

The abundance pattern for this model is shown in the 
right panel of Figure~\ref{fig:nucmod} as the blue stars.
Qualitatively, the match to the data is poorer
than in the previous case, with both C and N appearing
somewhat discrepant.
However, we caution against reading too much
into such a comparison: the abundances
of C and N have the largest errors among the 
elements we observed and this model involves
fewer free parameters.
Statistically, the two models shown in 
Figure~\ref{fig:nucmod} 
are equally probable, \textit{if 
the (unknown) model uncertainties
are comparable in the two cases}.

To determine which of the two model
suites provide a closer representation of
the data we use the Bayes evidence ratio
(otherwise known as the Bayes factor).
This measure has the advantage of being
able to directly compare two model suites
with different physical prescriptions and
a different number of free parameters. Using
the Laplace approximation (see e.g. \citealt{Mac03}, p. 341)
and assuming flat priors, we measure the evidence
ratio of the \citet{ChiLim04} models relative to
the \citet{HegWoo10} models to be ${\cal E}\sim1.5$,
confirming our earlier conclusion that it is 
difficult to discern between the two models.

In the right panel of Figure~\ref{fig:nucmod},
we also show the best-fitting model from the
recent \citet{LimChi12} metal-free nucleosynthesis
calculations (orange stars). These new models were
computed with the latest version of their stellar
evolutionary code, and now include higher mass stars
(with a total mass range of $13 M_{\odot}$ to $80 M_{\odot}$).
The best-fitting model with this suite of calculations is for
a $35 M_{\odot}$ metal-free star that ejects
$0.139\,$M$_{\odot}$ of $^{56}$Ni. 
Although the $Z=0.05\,\,Z_{\odot}$ model gives a somewhat
better fit, we consider both models to be adequate
representations of the data given the unknown model
uncertainties.

We have not considered models of rotating
low metallicity Population II \citep{MeyEksMae06,Hir07}
or Population III stars \citep{Jog10}.
These models tend to yield a large 
[N/Fe] abundance, something that we 
do not observe in the abundance pattern
of this DLA. We also refrain from computing
full-scale galactic chemical evolution models.
Such an exercise was performed recently by
\citet{SalFer12} to infer the enrichment history
of CEMP DLAs. These authors conclude that such DLAs
are enriched by Population III stars, but may receive
an additional contribution from AGB stars -- which
would increase both the C and N abundances (see e.g.
Table 6 from \citealt{Cri11}). Given the
relatively low N abundance observed here, we suggest
that the metal contribution from AGB stars for this
system is minimal.

Thus, on the basis of the available data,
it is not possible to distinguish statistically
between the two possibilities considered here,
enrichment solely by metal-free stars,
or by stars of moderate metallicity, with 
$Z \sim 1/20 Z_\odot$  (and of course
it would be even more
difficult to recognise a mixed scenario, involving
more than one stellar generation).
Looking ahead, there are two
foreseeable avenues that may yield tighter
constraints on the nature of the stars
that enrich the most metal-poor DLAs:

(1) Improvements in the modelling of
massive star nucleosynthesis and, in 
particular, a better description of
the physics behind the explosion
mechanism. Much progress is currently 
being made on the relevant theory
(see e.g. \citealt{Bel11}). 
Furthermore, future increases in 
computational power will allow the 
construction of large databases of 
finer-grid nucleosynthesis calculations.
This may remove the `discreteness' of
the current models.

(2) Stronger discriminatory power between
different models, particularly
regarding the explosion energy,
lies with 
the ratio of two Fe-peak elements
(e.g. [Cr,Ni,Zn/Fe], \citealt{UmeNom02}). 
In the most metal-poor DLAs, 
absorption lines from these less abundant
elements are too weak to be detected 
with current observational facilities,
but may be within the grasp of the
next generation of 30-m class telescopes.
This will open up new avenues to probe the 
details of the explosion.
Furthermore, the greater light gathering power
of these future facilities will make it
feasible to observe the faint background 
quasars with the spectral resolution 
$R\gtrsim 100\,000$ required to fully resolved
the metal absorption lines of most DLAs,
and thus decouple directly (rather than by indirect means
as done here) the relative contributions to the
line widths from 
thermal and turbulent broadening. 

\subsection{Identifying more CEMP DLAs}

At present, the discriminatory power 
between Population II and Population III 
models rests with the [C/Fe] ratio.
We should therefore compile a list of 
candidate systems, such as the one described
here, that exhibit the following hallmark 
signatures:
(1) an enhanced [C/Fe] ratio with relatively
normal [N/Fe] (i.e. $\lesssim0.0$). Current 
models of Population II nucleosynthesis have 
difficulty in reproducing these signatures. 
(2) an [O/Fe] ratio
that is enhanced beyond the typical value 
observed for DLAs ($\sim+0.4$; 
\citealt{Coo11b}); enhanced [C/Fe] is more
likely to yield enhanced [O/Fe] as well
(assuming the Population III nucleosynthesis 
calculations by \citealt{HegWoo10}).
Furthermore, there are several weak \OI\
lines that are insensitive to the details of 
the cloud model and in particular to the 
balance between thermal and turbulent 
broadening. The more easily measured 
O/Fe ratio may therefore 
prove to be a reliable indicator of further
examples of CEMP DLAs.

\section{Summary and Conclusions}
\label{sec:conc}
 
We have reported new observations,
at a higher spectral resolution than any
published previously,
of the very metal-poor damped \Lya\ system at
$z_{\rm abs}=3.067$ towards the QSO
J1358$+$6522.
Our Keck HIRES spectrum covers 21 absorption
lines from ion stages of H, C, N, O, Al, Si, S, and Fe
which are dominant in neutral gas.
From the analysis of these data we
draw the following conclusions:\\

\noindent ~~(i) The metallicity
of the DLA, as measured by the 
Fe/H ratio, is $\sim 1/700$ of solar
([Fe/H]\,$ = -2.84$).
Both C and O appear to be enhanced
by a factor of $\sim 4$ relative to Fe.

\smallskip

\noindent ~~(ii) 
The gas is `cold', in the sense that the
profiles of the absorption
lines are best reproduced by a model that 
has a low value of the Doppler parameter,
$b_{\rm tot} = 3.0$\,km\,s$^{-1}$,
which furthermore appears to be dominated by 
turbulent, rather than thermal, motions.

\smallskip

\noindent ~~(iii) We derived a $2\sigma$
upper limit on the DLA's kinetic temperature
of $T_{\rm DLA}\le 4700$\,K. 
The thermal broadening that
corresponds to this upper limit lowers
the [C/Fe] ratio by 0.37 dex,
but does not affect the [O/Fe] which
is deduced from weak, unsaturated absorption lines.

\smallskip

\noindent ~~(iv) Finally, we compared the
abundance pattern for this candidate CEMP DLA 
to model nucleosynthesis calculations for 
Population I, II, and III stars.
Given the handful of elements available
and the present limitations of the models,
we are unable to formally distinguish between models
of Population II and Population III enrichment.
Upcoming facilities will permit the abundances
for several Fe-peak elements to be measured,
providing additional diagnostics 
between these two possibilities.

CEMP DLAs might be more common than
currently appreciated; the available \CII\
lines are almost always saturated, making
it difficult to measure accurately the C 
abundance \emph{unless} one can determine a 
secure cloud model from a careful analysis of the 
line profiles. Given the 
potential of CEMP DLAs for probing 
nucleosynthesis by 
Population III stars, a sample of 
the most promising candidates should now
be compiled. With the next
generation of 30-metre class telescopes,
it should become feasible to record absorption
lines from the intrinsically less abundant 
Fe-peak elements, and thereby
explore in more detail than is possible at 
present the properties of the core-collapse 
supernova explosions of metal-free stars.

\section*{Acknowledgements}
We are grateful to 
the staff astronomers at the Keck  
Observatory for their expert assistance 
with the observations.
We also thank an anonymous referee who
provided some comments on the manuscript.
Valuable advice and help with various 
aspects of the work described in this 
paper was provided by Bob Carswell, and
Sergey Koposov. We thank the Hawaiian
people for the opportunity to observe from Mauna Kea;
without their hospitality, this work would not have been possible.
RC is supported by a Research Fellowship 
at Peterhouse College, Cambridge. 
MTM thanks the Australian Research Council 
for a QEII Research Fellowship (DP0877998).


\label{lastpage}

\end{document}